\documentclass[twocolumn]{aastex63}

\shorttitle{HAWC~J1825-134 as hadronic PeVatron}
\shortauthors{Dzhatdoev, Podlesnyi, and Vaiman}

\begin{document}
\title{The primary proton spectrum of the hadronic PeVatron candidate HAWC~J1825-134}
\correspondingauthor{Timur Dzhatdoev}
\email{timur1606@gmail.com}
\correspondingauthor{Egor Podlesnyi}
\email{podlesnyi.ei14@physics.msu.ru}
\author[0000-0002-7660-4236]{Timur Dzhatdoev}
\affiliation{Federal State Budget Educational Institution of Higher Education, M.V. Lomonosov Moscow State University, Skobeltsyn Institute of Nuclear Physics (SINP MSU), 1(2), Leninskie gory, GSP-1, 119991 Moscow, Russia}
\affiliation{Institute for Nuclear Research of the Russian Academy of Sciences, 60th October Anniversary Prospect 7a, Moscow 117312, Russia}
\affiliation{Institute for Cosmic Ray Research, University of Tokyo, 5-1-5 Kashiwanoha, Kashiwa, Japan}
\author[0000-0003-3395-0419]{Egor Podlesnyi}
\affiliation{Federal State Budget Educational Institution of Higher Education, M.V. Lomonosov Moscow State University, Department of Physics, 1(2), Leninskie gory, GSP-1, 119991 Moscow, Russia}
\affiliation{Federal State Budget Educational Institution of Higher Education, M.V. Lomonosov Moscow State University, Skobeltsyn Institute of Nuclear Physics (SINP MSU), 1(2), Leninskie gory, GSP-1, 119991 Moscow, Russia}
\affiliation{Institute for Nuclear Research of the Russian Academy of Sciences, 60th October Anniversary Prospect 7a, Moscow 117312, Russia}
\author[0000-0002-8255-3631]{Igor Vaiman}
\affiliation{Federal State Budget Educational Institution of Higher Education, M.V. Lomonosov Moscow State University, Department of Physics, 1(2), Leninskie gory, GSP-1, 119991 Moscow, Russia}
\affiliation{Federal State Budget Educational Institution of Higher Education, M.V. Lomonosov Moscow State University, Skobeltsyn Institute of Nuclear Physics (SINP MSU), 1(2), Leninskie gory, GSP-1, 119991 Moscow, Russia}

\begin{abstract}
The $\gamma$-ray spectrum of the source HAWC~J1825-134 measured with the High Altitude Water Cherenkov (HAWC) observatory extends beyond 200 TeV without any evidence for a steepening or cutoff. There are some indications that the $\gamma$-rays detected with HAWC were produced by cosmic ray protons or nuclei colliding with the ambient gas. Assuming primary protons, we inquire which shape of the primary proton spectrum is compatible with the HAWC measurements. We find that the primary proton spectrum with the power-law shape of $\gamma_{p} = 2.2$ and the cutoff energy $E_{c-p} > 500$ TeV describes the data well. However, much harder spectra with $\gamma_{p}$ down to 1.3 and $E_{c-p}$ as low as 200 TeV also do not contradict the HAWC measurements. The former option might be realized if the accelerator is inside or very near to the $\gamma$-ray production zone. The latter option is viable for the case of a cosmic ray source which effectively confines low-energy ($E_{p} < 10$ TeV) accelerated protons. Using publicly-available data of the Fermi-LAT space $\gamma$-ray telescope, we derive upper limits on the intensity of the \object{HAWC~J1825-134} source in the 1 GeV -- 1 TeV energy range. We show that the account of these upper limits drastically changes the interpretation: only hard ($\gamma_{p} < 1.7$) spectra describe the combined HAWC and Fermi-LAT datasets well.
\end{abstract}
\keywords{\textit{Unified Astronomy Thesaurus concepts:} Gamma-ray astronomy (628); Gamma-ray sources (633); Gamma-rays (637); Gamma-ray observatories (632); Galactic cosmic rays (567); High energy astrophysics (739)}

\section{Introduction} \label{sec:introduction}

Cosmic rays up to the so-called ``knee'' --- a steepening in the all-nuclei  spectrum measured at Earth \citep{Kulikov1959,Aglietta2004,Antoni2005} --- are widely believed to have a Galactic origin. The acceleration and propagation of cosmic ray (CR) protons and nuclei colliding with the ambient gas is accompanied by the production of $\gamma$-rays \citep{Stecker1970,Stecker1973,Black1973,Caraveo1980,Ackermann2012a,Albert2021a}. Therefore, Galactic hadronic PeVatrons --- the objects accelerating protons up to the knee --- could be searched for and studied with $\gamma$-astronomical methods.

So far, the search for these objects was not very successful. Indeed, $\gamma$-ray spectra of supernova remnants reveal a high-energy cutoff \citep{Ahnen2017,Abdalla2018}. The spectrum of the Cygnus Cocoon complex containing star-forming regions that could accelerate cosmic rays \citep{Montmerle1979,Cesarsky1983,Bykov2014,Bykov2018,Aharonian2019} also reveals a steepening at 10~TeV \citep{Abeysekara2021}.\footnote{however, the authors argue that the time-averaged CR spectrum in some time-dependent models could be harder than in stationary models} The superhigh energy ($E > 100$~TeV)  emission from pulsar wind nebulae \citep{Amenomori2019,Abeysekara2020} is well explained with the leptonic mechanism \citep{Khangulyan2019,Breuhaus2021}. PeV protons could be accelerated near the Galactic Center \citep{Abramowski2016}, but it is not clear whether these protons could make a substantial contribution to the CR flux observed at Earth.

Very recently, the High Altitude Water Cherenkov (HAWC) collaboration reported a measurement of the $\gamma$-ray spectrum of the source \object{HAWC~J1825-134} up to the energy of 300 TeV (\citet{Albert2021b}, hereafter A21). The spectrum does not reveal any steepening or cutoff. A21 argue that the observed $\gamma$-rays were produced by the hadronic mechanism.

In the present paper we put constraints on the parameters of the primary CR spectrum, taking the HAWC results at face value and assuming that all $\gamma$-rays detected by HAWC from this source were produced by primary protons.\footnote{more precisely, by secondary particles produced by primary protons} In Section~\ref{sec:region}, we dissect the region of the sky containing \object{HAWC~J1825-134}. In Section~\ref{sec:Fermi}, we set upper limits on the spectral energy distribution (SED~=~$E_{\gamma}^{2}dN_{\gamma}/dE_{\gamma}$) of the source in the energy range of 1 GeV -- 1 TeV using publicly-available Fermi Large Area Telescope (Fermi-LAT) data  \citep{Atwood2009}.

In Section~\ref{sec:examples}, we provide a few examples of a plausible primary proton spectrum that could explain the HAWC data reasonably well. We perform a scan on the parameters of the primary proton spectrum in Section~\ref{sec:scan}. Finally, we discuss the obtained results in Section~\ref{sec:discussion} and conclude in Section~\ref{sec:conclusions}.

\section{The HAWC~J1825-134 source \\ and its surroundings \label{sec:region}}

\begin{figure*}[t!]
\includegraphics[width=1.0\textwidth]{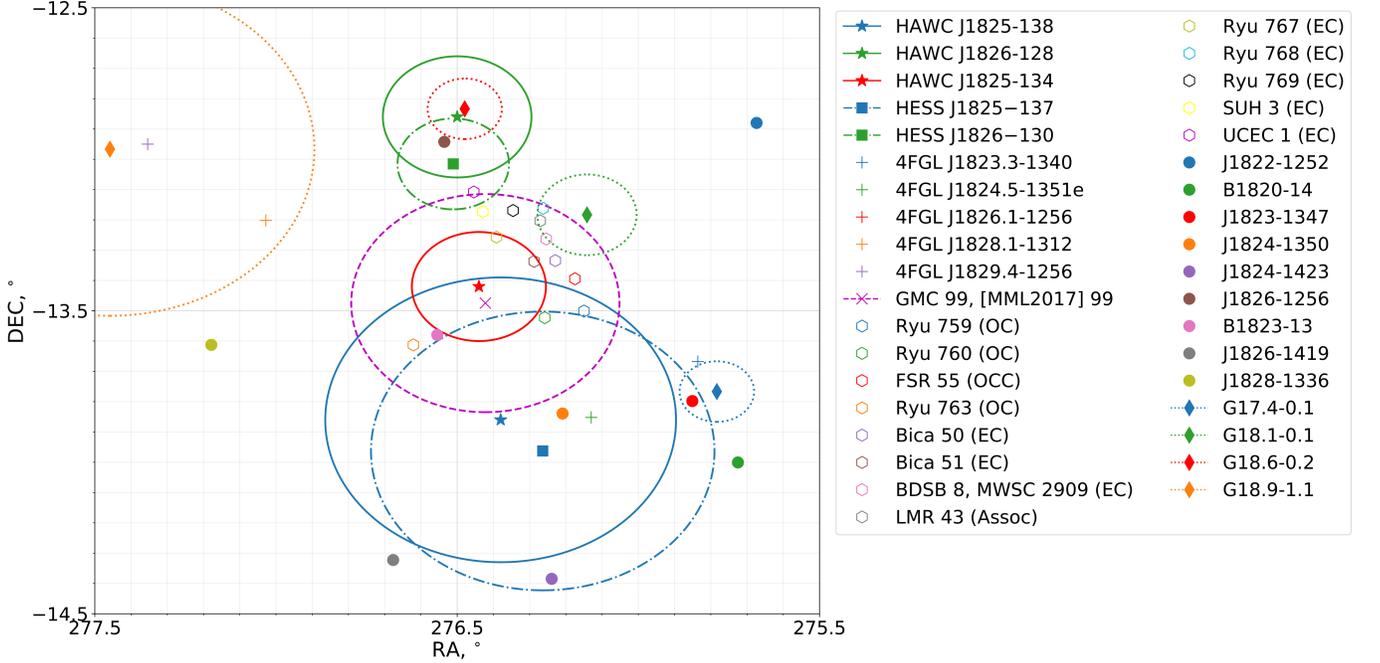}
\caption{A view of the HAWC~J1825-134 region (see the text for more details).} \label{Fig-View}
\end{figure*}

The \object{HAWC~J1825-134} source is situated in a crowded region of the $\gamma$-ray sky (see Fig.~\ref{Fig-View}). In A21, the HAWC collaboration presents a new analysis of this region, identifying three bright TeV sources\footnote{besides the Galactic diffuse emission} shown in Fig.~\ref{Fig-View} as stars: \object{HAWC~J1825-134} itself, as well as \object{HAWC J1826-128} and \object{HAWC J1825-138}. The H.E.S.S. imaging Cherenkov telescope array (IACT) detected the likely counterparts of \object{HAWC J1826-128} and \object{HAWC J1825-138} --- the sources \object{HESS J1826-130} \citep{Abdalla2020} and \object{HESS J1825-137} \citep{Abdalla2019}, respectively (denoted by squares in the figure). The extensions for HAWC sources are shown as solid circles\footnote{for \object{HAWC J1825-138} and \object{HAWC J1826-128} circles represent widths from Table 1 in A21, and for \object{HAWC J1825-134} the circle represents the $0.18^{\circ}$ upper limit on the extension of the source (95\% confidence level)}, for H.E.S.S. sources --- as dot-dashed circles. Statistical and systematic uncertainties of H.E.S.S. and HAWC source position measurements are not shown. The positions of Fermi-LAT sources from the 4FGL catalog \citep{Abdollahi2020} (without extensions and position measurement uncertainties) are shown in Fig.~\ref{Fig-View} as plus signs.

The giant molecular cloud \object{GMC 99} (an alternative name is \object{[MML2017] 99}) \citep{Miville-Deschnes2016} with the angular radius of $0.36^{\circ}$ is shown in Fig. \ref{Fig-View} as dashed magenta circle; its center is denoted as magenta cross. Hollow hexagons denote objects from the multi-band catalog of star clusters, associations and candidates according to \citet{Bica2018} (angular sizes are not shown and only objects within an $0.4^{\circ}$-square with the center at (276.5$^{\circ}$, -13.5$^{\circ}$) are shown). Filled circles denote pulsars from the ATNF catalog of \citet{Manchester2005, ATNF}; diamonds --- SNRs with their extensions shown as dotted circles according to \citet{Green2019, Green}. Object names are shown in the legend, for the objects from the catalog of \citet{Bica2018} the object type is shown in brackets after its name(s): OC~--- open cluster, OCC~--- open cluster candidate, EC~--- embedded cluster, Assoc~--- star association.

Star-forming regions are among the possible sources of Galactic cosmic rays up to the knee. In particular, primary protons producing $\gamma$-rays responsible for the \object{HAWC~J1825-134} source could have been accelerated in the embedded young star cluster \object{BDSB~8} (\object{MWSC~2909}) as proposed in A21 (the distance $D~\approx~4.4$~kpc from the Sun, \citep{Kharchenko2015, Kharchenko2015b}. A tentative target for protons producing $\gamma$-ray emission of the \object{HAWC~J1825-134} source is the core of \object{GMC~99} ($D~\approx~3.9$~kpc \citep{Miville-Deschnes2016}). For \object{Bica~51} $D~\approx~1.4$~kpc \citep{Kharchenko2015}, for \object{SUH~3} $D~\approx~2.9$~kpc \citep{Kurtz1994}; therefore, these are unlikely sources of protons in \object{HAWC~J1825-134}. For \object{LMR~43}, $D~\approx 3.7~\pm~1.7$~kpc \citep{Lee2012}; for ECs \object{Bica~50}, \object{Ryu~767}, \object{Ryu~768}, \object{Ryu~769}, \object{UCEC~1} we could not find estimates of $D$. Therefore, we cannot exclude them from being possible sources of very high-energy protons for \object{HAWC~J1825-134}.

\section{Fermi-LAT data analysis}  \label{sec:Fermi}

\begin{figure*}
    \vspace{1pc}
    \includegraphics[width=0.50\textwidth]{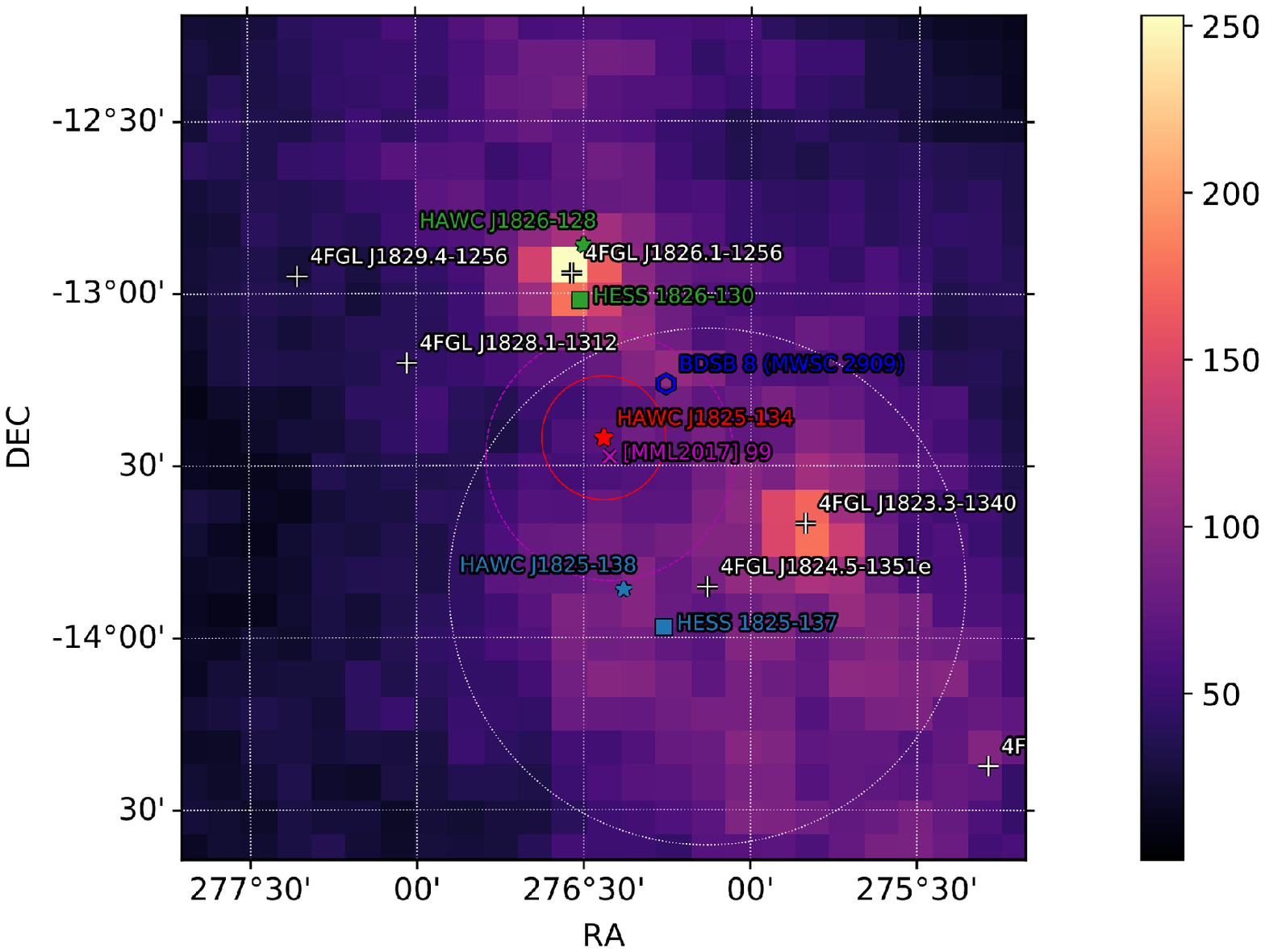}
    \includegraphics[width=0.50\textwidth]{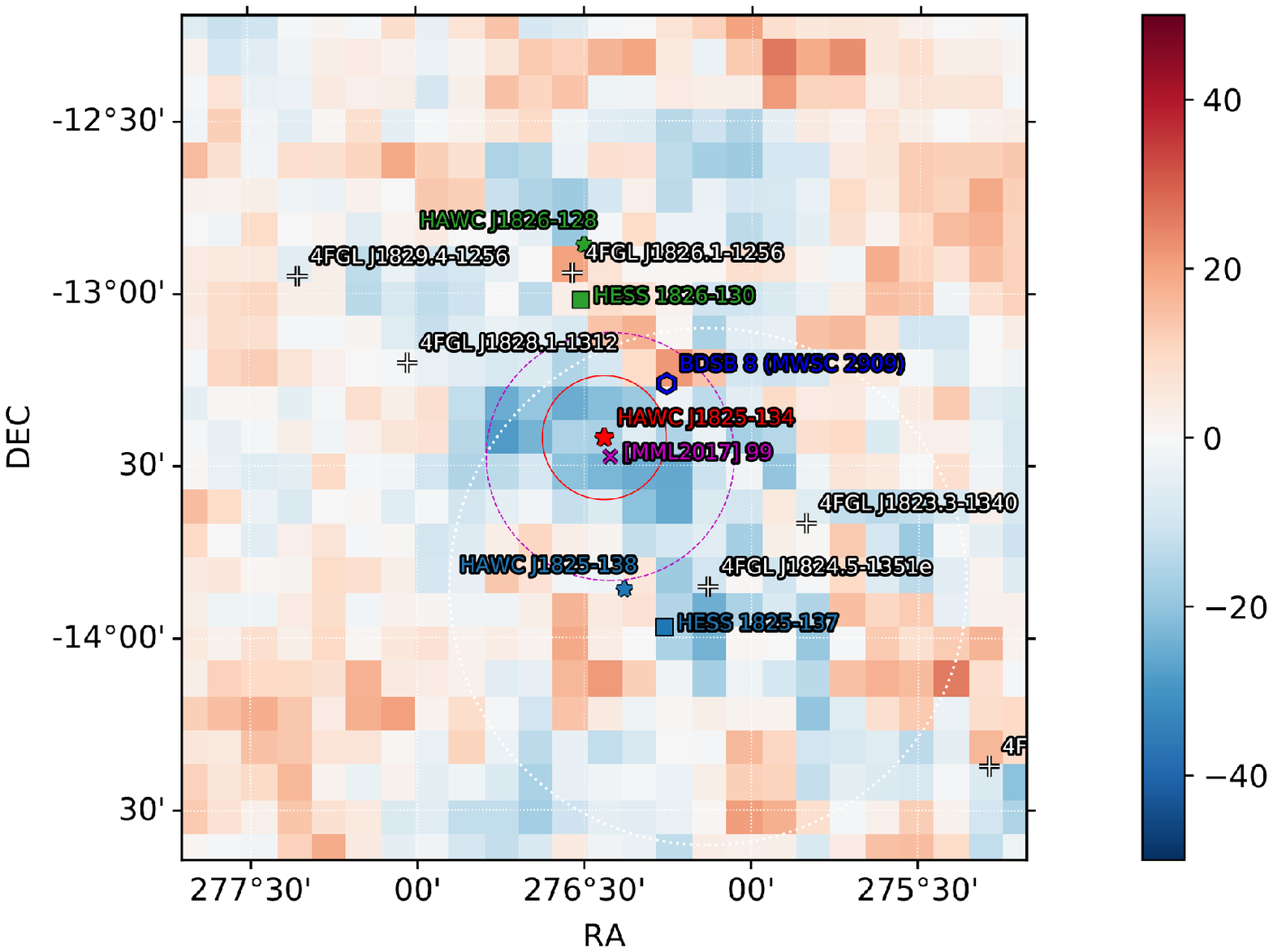}
\caption{Fermi-LAT maps ($E \geq 10$~GeV) of the \object{HAWC~J1825-134} region. Left: Fermi-LAT count map. Right: Fermi-LAT excess map.\label{Fig-View2}}
\end{figure*}

Remarkably, the \object{HAWC~J1825-134} source does not have any 4FGL counterpart. We use Fermi-LAT data\footnote{\url{https://fermi.gsfc.nasa.gov/cgi-bin/ssc/LAT/LATDataQuery.cgi}} from 4$^{\mathrm{th}}$ of August, 2008 to 28$^{\mathrm{th}}$ of January, 2021 in the energy range from 1 GeV to 1 TeV. The region of interest (ROI) is a circle with the radius $R_{\mathrm{ROI}} = 10^{\circ}$ centered at the best fit position of \object{HAWC~J1825-134} $(276.44^{\circ}, -13.42^{\circ})$ reported in A21.

We performed a binned likelihood analysis of this dataset using \textit{fermipy} \citep{Wood2017} (version 0.17.4) and \textit{fermitools}\footnote{\url{https://github.com/fermi-lat/Fermitools-conda}} (version 1.2.23) packages assuming a model of the ROI containing the sources from the 4FGL catalog \citep{Abdollahi2020}, the diffuse galactic background, the isotropic $\gamma$-ray background and the model of the source in question as a point source with a simple power law spectrum at the center of the ROI. Some details of this analysis are available in Appendix \ref{Appendix:analysis_details}.

The test statistic $TS$ corresponding to the hypothesis of the source \object{HAWC~J1825-134} being present in the model against the null hypothesis of the source being absent was calculated. We obtained $TS \ll 1$, i.e. no significant signal from the source was observed.

A21 note that there might be a systematic shift about $0.2^{\circ}$ in the \object{HAWC~J1825-134} source position. To ensure that this systematic error does not modify the negative result of our search we also performed the source localization procedure using the \texttt{fermipy.GTAnalysis.localize} method searching for a better position of the source in question inside the $0.5^{\circ}$ square with the center coinciding with the ROI center. This search yielded negative results.

Fig. \ref{Fig-View2} shows the surroundings of \object{HAWC~J1825-134} as seen by Fermi-LAT for the energy $E \geq 10$~GeV\footnote{we show maps for $E \geq 10$~GeV since Fermi-LAT reaches the best angular resolution about $0.1^{\circ}$ at energies $E \geq 10$~GeV allowing to better discern different sources}: the count map with the pixel size of $0.1^{\circ}$ per pixel is shown on the left, and the excess count map (i.e. the map of the difference between the observed number of counts and the predicted number of counts of the fitted model in every pixel) with the same pixel size is shown on the right.

The position of \object{HAWC~J1825-134} is marked with red star with red solid circle representing the upper limit on its extension. The H.E.S.S. sources are shown as green and light blue squares, other HAWC sources~--- as  green and light blue stars, \object{[MML2017] 99}~--- as magenta cross with the extension shown as magenta dashed circle, EC \object{BDSB 8} (\object{MWSC 2909})~--- as blue hollow hexagon, Fermi-LAT sources~--- as white plus signs.

The model describes the data reasonably well. Although there is some excess in some pixels near the pulsar \object{4FGL J1826.1-1256}, this excess overshoots the model by only about $10\%$ and occurs far from the position of \object{HAWC~J1825-134} reported in A21. Since for the \object{HAWC~J1825-134} source $TS \ll 1$, it does not yield any predicted counts in the best fit model. In the $0.18^{\circ}$-vicinity of \object{HAWC J1825-134} there is even a slight deficit of photon counts. The source in question is located inside the extension circle of the pulsar wind nebula (PWN) \object{4FGL J1824.5-1351e} with the extension radius $0.75^{\circ}$ \citep{Abdollahi2020} shown in Fig. \ref{Fig-View2} as white dotted circle.

Comprehensive studies of this PWN associated with \object{HESS J1825-137} were carried out by \citet{Araya2019} and \citet{Principe2020}, and they did not find a source coincident with the position of \object{HAWC~J1825-134}. Thus our negative result of the search for the $\gamma$-ray emission from \object{HAWC~J1825-134} is in agreement with the previous studies.

\begin{figure}[ht!]
\vspace{1pc}
\includegraphics[width=8.5cm]{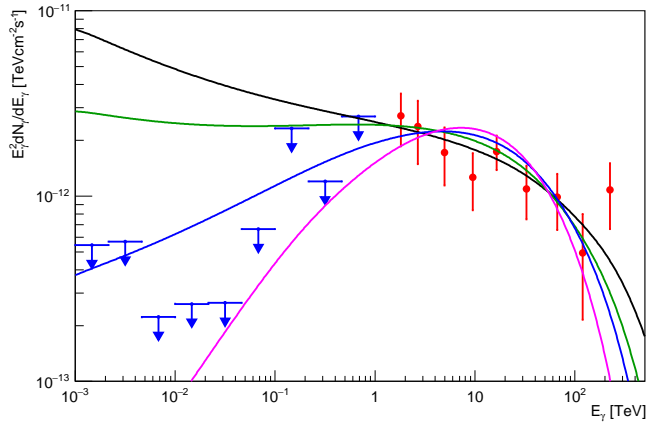}
\caption{SED of HAWC~J1825-134 measured with HAWC (red circles with statistical uncertainties) together with Fermi-LAT upper limits (blue arrows) as well as four model curves calculated for various proton spectrum parameters (see text for more details). \label{Fig-Spectrum}}
\end{figure}

\begin{figure*}[ht!]
    \vspace{1pc}
    \includegraphics[width=0.50\textwidth]{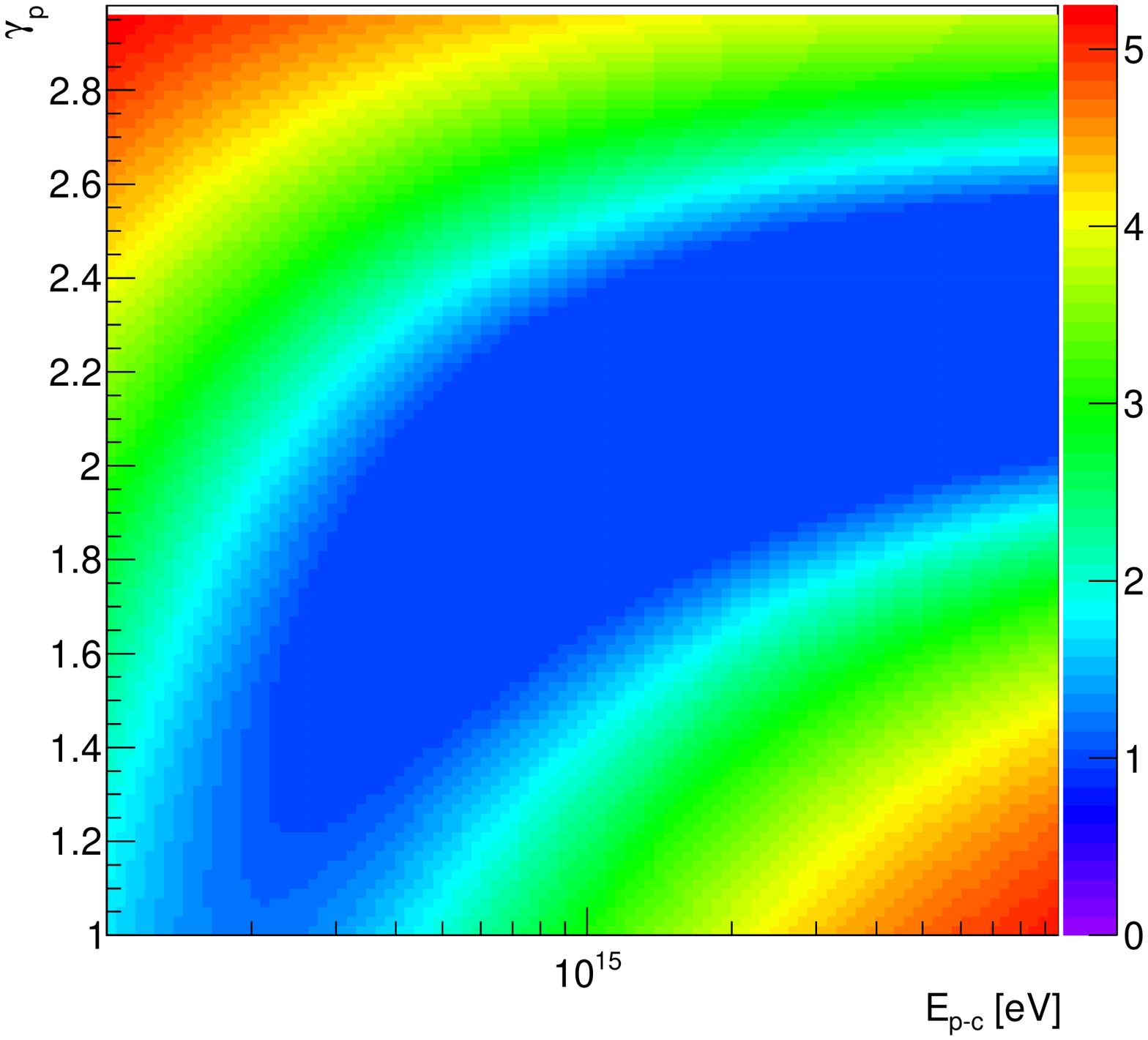}
    \includegraphics[width=0.50\textwidth]{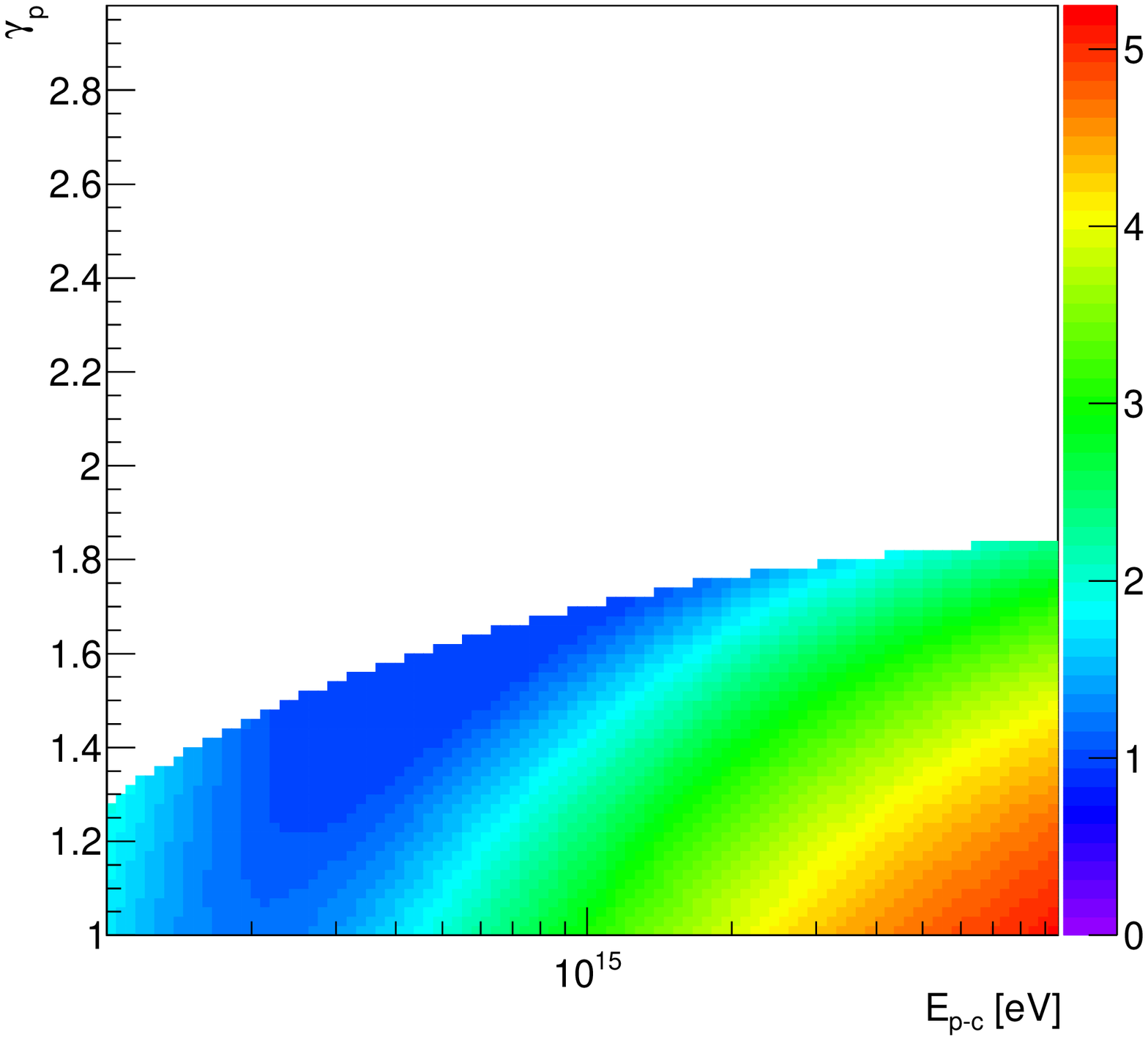}
\caption{Exclusion significance (Z-value) vs. the primary proton spectrum parameters. Left: only HAWC data accounted for. Right: HAWC data and Fermi-LAT upper limits accounted for (the Fermi-LAT upper limits are overshot in the white region of the parameter area). \label{Fig-Significance}}
\end{figure*}

Finally, we obtained upper limits (95\% confidence level) on the SED of \object{HAWC~J1825-134} using \texttt{fermipy.GTAnalysis.sed} method allowing the source in question to have different values of spectral index in different energy bins. These upper limits are shown in Fig.~\ref{Fig-Spectrum} as blue arrows. In Appendix \ref{Appendix:analysis_extended} we compare the upper limits on the SED of \object{HAWC~J1825-134} under the point-like source and extended source hypotheses.

\section{Examples of the primary proton spectrum} \label{sec:examples}

In what follows we take the HAWC results concerning the analysis of the HAWC~J1825-134 source at face value. The spectral energy distribution of HAWC~J1825-134 according to A21 is shown in Figure~\ref{Fig-Spectrum} as red circles. Four model $\gamma$-ray SEDs are also shown in this Figure, assuming a hadronuclear (more precisely, proton-proton) mechanism of $\gamma$-ray production on optically thin matter.\footnote{that is, neglecting the contribution of cascade $\gamma$-rays to the observable $\gamma$-ray spectrum; given that the average gas column density from the region of HAWC~J1825-134 is $N_{H} = 3\times10^{22}$ cm$^{-2}$ (see A21), this assumption is justified}

These SEDs were calculated using approximations of \citet{Kelner2006} under the assumption of the primary proton spectrum $\propto E^{-\gamma_{p}}exp(-E_{p}/E_{p-c})$. Black curve corresponds to the case of $\gamma_{p} = 2.2$ and $E_{p-c} = 3$~PeV, green curve~--- $\gamma_{p} = 2.06$ and $E_{p-c} = 1$ PeV, blue curve~--- $\gamma_{p} = 1.8$ and $E_{p-c} = 500$ TeV, magenta curve~--- $\gamma_{p} = 1.3$ and $E_{p-c} = 200$ TeV.

For simplicity, we neglect bremsstrahlung of electrons produced together with $\gamma$-rays. The $\gamma$-ray absorption effect in the Galactic volume between the source and the observer was included according to \citet{Vernetto2016}. Since this effect is small, our results do not change if other model (e.g. \citet{Moskalenko2006}) is chosen.\footnote{in particular, these models account for the absorption on cosmic microwave background (CMB) photons} We leave the intrinsic absorption effect for future studies.

\section{Scan over parameters} \label{sec:scan}

We performed an exhaustive scan over the values of the primary proton spectrum parameters $(\gamma_{p}, E_{p-c})$ in the range $\gamma_{p} = 1-3$ and $E_{p-c} = 10^{14}-10^{16}$ eV. For every set of the parameters, a model histogram of the observable SED was computed in the HAWC energy bins. After that, the optimal normalization of the model SED was determined that minimizes the value of the $\chi^{2}$ form. The optimized $\chi^{2}$ value was converted to p-value and then to Z-value (statistical significance) according to the prescriptions of \citet{Zyla2020}. The resulting 100$\times$100 matrix of the Z-value vs. $(\gamma_{p}, E_{p-c})$ is shown in Figure~\ref{Fig-Significance} (left). All values of the Z-value below 1.0 are shown in Figure~\ref{Fig-Significance} (left) in the same color.

In particular, it turns out that the primary proton spectrum with parameters $\gamma_{p} = 2.2$ and $E_{p-c} = 3$ PeV describes the HAWC data well. However, some harder spectra with relatively low values of $E_{p-c}$ (such as $\gamma_{p} = 2.06$ and $E_{p-c} = 1$ PeV; $\gamma_{p} = 1.8$ and $E_{p-c} = 500$ TeV; or even $\gamma_{p} = 1.3$ and $E_{p-c} = 200$ TeV) formally do not contradict the HAWC data.

In Figure~\ref{Fig-Significance} (right) we show the result of the same procedure, but accounting for the Fermi-LAT upper limits. The area contradicting these upper limits is shown in white color. We conclude that the account of the Fermi-LAT upper limits drastically changes the interpretation, allowing to exclude soft ($\gamma_{p} > 1.8$) primary proton spectra for all considered values of $E_{p-c}$.

The assumption of the log-parabolic  or broken power-law primary proton spectrum does not change our conclusions qualitatively. Finally, we note that the account of an additional $\gamma$-ray component which contributes mainly at relatively low energies (for instance, $\gamma$-rays from bremsstrahlung) would require even harder primary proton spectra, reinforcing our conclusions.

\section{Discussion} \label{sec:discussion}

\subsection{The acceleration spectrum and the effective $\gamma$-ray production spectrum}

It is widely believed that a typical Galactic hadronic PeVatron has a spectrum of protons with $\gamma_{p} = 2.0-2.3$ and $E_{p-c} = 1-3$ PeV. However, the $\gamma$-ray production zone is not necessarily spatially coincident with the particle acceleration zone. The primary proton spectrum may be significantly modified by propagation effects \citep{Aharonian1996} associated with escape from the accelerator (e.g. \citet{Moskalenko2008}), diffusion from the accelerator to the $\gamma$-ray production zone, etc. Therefore, a hard ``effective'' spectrum of protons with $\gamma_{p} = 1.8$ or even $\gamma_{p} = 1.3$ could not be excluded a priori. We note that \citet{Bykov2018} predicted very hard spectra of accelerated particles at the energy significantly below the maximum acceleration energy; our results are consistent with this prediction.

\subsection{Statistical considerations}

The results of this study could be somewhat influenced by several statistical effects, including a) migration of events between the energy bins of the HAWC spectrum, b) a possible Eddington bias in the last energy bin. An account of the first effect is likely to result in an increase of the minimal value of $E_{p-c}$ compatible with the HAWC data. The Eddington bias would lead to an overestimation of the observable flux at the last energy bin. Therefore, a correction for this statistical effect would decrease the minimal value of $E_{p-c}$ compatible with the HAWC data. To some extent, these two effects partially compensate each other.

\subsection{Prospects of neutrino detection from HAWC~J1825-134}

\object{HAWC~J1825-134} is a promising Galactic neutrino source \citep{Niro2019}. Therefore, some hints about the nature of this source could potentially be obtained from astrophysical neutrino observations. We leave this study for future works.

\section{Conclusions} \label{sec:conclusions}

The HAWC data alone do not allow one to exclude the hypothesis of $E_{p-c} < 1$ PeV for the \object{HAWC~J1825-134} source. Formally, a wide range of $(\gamma_{p}, E_{p-c})$ (down to $\gamma_{p} = 1.3$ and $E_{p-c} = 200$ TeV) do not contradict the spectrum of this source measured with HAWC. The account of the Fermi-LAT upper limits allowed us to exclude soft ($\gamma_{p} > 1.8$) primary proton spectra for all considered values of $E_{p-c}$. Observations with the LHAASO \citep{LHAASO-arxiv-2019} and CTA \citep{Actis2011} detectors will be crucial in establishing the nature of this source.

\acknowledgments

We are grateful to Prof. P. Lipari and Dr. S. Vernetto for sharing their model of $\gamma$-ray absorption in the Galaxy \citep{Vernetto2016}. The authors acknowledge helpful discussions with Prof. S.V. Troitsky and Prof. I.V. Moskalenko. This work is supported in the framework of the State project ``Science'' by the Ministry of Science and Higher Education of the Russian Federation under the contract 075-15-2020-778. E. P. thanks the Foundation for the Advancement of Theoretical Physics and Mathematics ``BASIS'' (Contract No. 20-2-10-7-1) and the Non-profit Foundation for the Development of Science and Education ``Intellect'' for the student scholarships.

\software{\textit{NumPy} \citep{Harris2020}, \textit{Astropy} \citep{Astropy2018}, \textit{Matplotlib} \citep{Hunter2007}, \textit{ROOT} \citep{Brun1997}.}

\facility{HAWC, Fermi-LAT.}

\newpage
\appendix
\section{Fermi-LAT data analysis details}
\label{Appendix:analysis_details}

In our Fermi-LAT data analysis of the $10^{\circ}$ ROI centered at the position of \object{HAWC J1825-134} we apply a cut on the zenith angle value $\theta_z \leq 90^{\circ}$ in order to remove contamination from the Earth's limb. We used the \texttt{{P8R3\_SOURCE\_V2}} instrument response function (\texttt{evclass}~$=128$) with events of both front and back type of the $\gamma$-ray conversion location (\texttt{evtype}~$=3$)\footnote{\url{https://fermi.gsfc.nasa.gov/ssc/data/analysis/documentation/Cicerone/Cicerone_LAT_IRFs/IRF_overview.html}}. To describe the observed $\gamma$-ray emission we constructed a model containing all sources from the 4$^{\mathrm{th}}$ Fermi-LAT source catalog 4FGL \citep{Abdollahi2020} located within $15^{\circ}$ from the center of the ROI, the model of the diffuse galactic background \texttt{gll\_iem\_v07}, the model of the isotropic $\gamma$-ray background \texttt{iso\_P8R3\_SOURCE\_V2\_v1}\footnote{\url{https://fermi.gsfc.nasa.gov/ssc/data/access/lat/BackgroundModels.html}} and the model of the source in question as a point source with a simple power law spectrum\footnote{\url{https://fermi.gsfc.nasa.gov/ssc/data/analysis/scitools/xml_model_defs.html}} at the center of the ROI. Spectral shapes and normalization of diffuse backgrounds and the source in question were left free, normalization of other sources within $5^{\circ}$ from the center of the ROI was also left free, but their spectral shapes were fixed; both normalization and spectral shapes of other sources outside $5^{\circ}$ from the ROI center were fixed to their catalog values. In our analysis the energy dispersion was taken into account for all sources except for the diffuse galactic background and the isotropic $\gamma$-ray background. Using this ROI model we performed the maximization of the binned likelihood function in the considered energy range using \texttt{fermipy.GTAnalysis.optimize} and \texttt{fermipy.GTAnalysis.fit} methods. The results of the likelihood analysis are discussed in Section \ref{sec:Fermi}.

\section{Analysis with the extended source model}
\label{Appendix:analysis_extended}

\begin{figure}[ht!]
\centering
\vspace{1pc}
\includegraphics[width=0.60\textwidth]{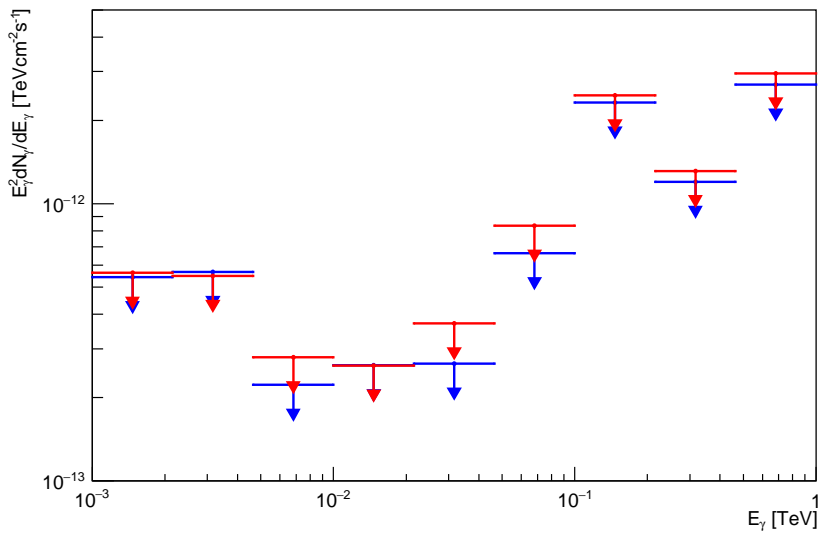}
\caption{Comparison of the upper limits on the SED of \object{HAWC J1825-134} for the point source (blue arrows) and extended source (red arrows) spatial models. \label{Fig-SED_extended}}
\end{figure}

We also performed a Fermi-LAT data analysis with the extended source model. All parameters of the constructed model were the same as described in Appendix \ref{Appendix:analysis_details}, except for the spatial model of \object{HAWC J1825-134}, which has been changed from the point source to the extended symmetric 2D radial Gaussian source with the $95\%$ containment radius of $0.18^{\circ}$, which is equal to the $95\%$ confidence level upper limit on the extension of \object{HAWC J1825-134} derived in A21. The obtained upper limits on the SED for both spatial models of \object{HAWC J1825-134} are shown in Fig.~\ref{Fig-SED_extended}. We note that the slight difference between the two versions of the SED upper limits does not modify our conclusions.

\bibliography{Pevatron-HAWC.bib}

\end{document}